\documentclass[iop,apjl]{emulateapj}
\usepackage{amsfonts}

\newcommand{\Lfir}{\mbox{$L_{\mathrm{FIR}}$}}

\newcommand{\MHtwo}{\mbox{$M_{\mathrm{H_{2}}}$}}
\newcommand{\Lsun}{\mbox{$L_{\sun}$}}
\newcommand{\Msun}{\mbox{$M_{\sun}$}}

\shorttitle{Far-Infrared Line Deficits}
\shortauthors{Graci{\'a}-Carpio et al.}

\begin{document}

\title{Far-Infrared Line Deficits in Galaxies with Extreme $\Lfir/\MHtwo$ Ratios\footnote{Herschel is an ESA space observatory with science instruments provided by European-led Principal Investigator consortia and with important participation from NASA.}}

\author{J. Graci{\'a}-Carpio\altaffilmark{1}, E. Sturm\altaffilmark{1}, S. Hailey-Dunsheath\altaffilmark{1}, J. Fischer\altaffilmark{2}, A. Contursi\altaffilmark{1}, A. Poglitsch\altaffilmark{1}, R. Genzel\altaffilmark{1}, E. Gonz{\'a}lez-Alfonso\altaffilmark{3}, A. Sternberg\altaffilmark{4}, A. Verma\altaffilmark{5}, N. Christopher\altaffilmark{5}, R. Davies\altaffilmark{1}, H. Feuchtgruber\altaffilmark{1}, J. A. de Jong\altaffilmark{1}, D. Lutz\altaffilmark{1}, L. J. Tacconi\altaffilmark{1}}

\altaffiltext{1}{Max-Planck-Institute for Extraterrestrial Physics (MPE), Giessenbachstra{\ss}e 1, 85748 Garching, Germany}
\altaffiltext{2}{Naval Research Laboratory, Remote Sensing Division, 4555 Overlook Ave SW, Washington, DC 20375, USA}
\altaffiltext{3}{Universidad de Alcal{\'a} de Henares, 28871 Alcal{\'a} de Henares, Madrid, Spain}
\altaffiltext{4}{Tel Aviv University, Sackler School of Physics \&\ Astronomy, Ramat Aviv 69978, Israel}
\altaffiltext{5}{Oxford University, Dept. of Astrophysics, Oxford OX1 3RH, UK}
\email{jgracia@mpe.mpg.de}

\begin{abstract}
We report initial results from the far-infrared fine structure line observations of a sample of 44 local starbursts, Seyfert galaxies and infrared luminous galaxies obtained with the PACS spectrometer on board \emph{Herschel}. We show that the ratio between the far-infrared luminosity and the molecular gas mass, $\Lfir/\MHtwo$, is a much better proxy for the relative brightness of the far-infrared lines than $\Lfir$ alone. Galaxies with high $\Lfir/\MHtwo$ ratios tend to have weaker fine structure lines relative to their far-infrared continuum than galaxies with $\Lfir/\MHtwo \lesssim 80\,\Lsun\,\Msun^{-1}$. A deficit of the [\ion{C}{2}]\,158\,\micron\ line relative to \Lfir\ was previously found with the ISO satellite, but now we show for the first time that this is a general aspect of all far-infrared fine structure lines, regardless of their origin in the ionized or neutral phase of the interstellar medium. The $\Lfir/\MHtwo$ value where these line deficits start to manifest is similar to the limit that separates between the two modes of star formation recently found in galaxies on the basis of studies of their gas-star formation relations. Our finding that the properties of the interstellar medium are also significantly different in these regimes provides independent support for the different star forming relations in normal disk galaxies and major merger systems. We use the spectral synthesis code Cloudy to model the emission of the lines. The expected increase of the ionization parameter with $\Lfir/\MHtwo$ can simultaneously explain the line deficits in the [\ion{C}{2}], [\ion{N}{2}] and [\ion{O}{1}] lines. 
\end{abstract}


\section{Introduction}

\begin{figure*}
\centering
\scalebox{0.5}{\includegraphics{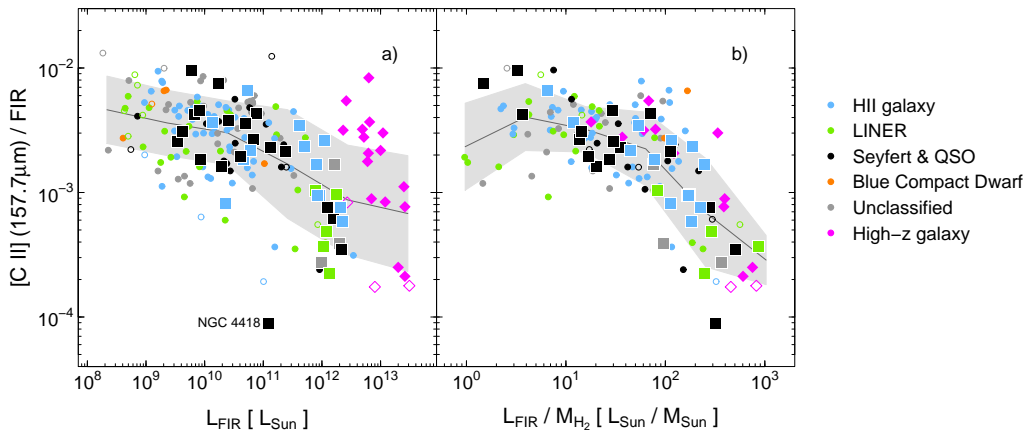}}
\caption{The global [\ion{C}{2}]\,158\,\micron\ line to FIR continuum ratio in galaxies with different optical activity classifications as a function of \Lfir\ (a) and $\Lfir/\MHtwo$ (b). PACS observations are represented with squares. Open symbols indicate 3$\sigma$ upper limits to the line flux. Grey areas connect points at one standard deviation below and above the mean (black line) for various \Lfir\ and $\Lfir/\MHtwo$ bins. \label{fig1}}
\end{figure*}

During most of their life, galaxies form stars at a relatively low rate, regulated by the total amount of molecular gas mass available \citep[e.g.,][]{Bigiel08}. This phase of semi-continuous star formation can be interrupted by short periods of violent star formation, normally associated with the merging of massive galaxies. In mergers most of the molecular gas is compressed on a scale of a few hundred parsec to a few kpc around the nuclei and, by a mechanism not yet well understood, forms stars with high efficiency \citep{Sanders96}. These episodes of intense star formation do not cover the whole merging process and might not be activated in all types of galaxy interactions, since many optically classified mergers do not show at present a significant increase in their star formation rates \citep[e.g.,][]{Bushouse88}.

The ratio between the far-infrared luminosity and the molecular gas mass, $\Lfir/\MHtwo$, is expected to be proportional to the number of stars formed in the galaxy per unit molecular gas mass and time (\MHtwo\ includes a 36\%\ mass correction for helium, and \Lfir\ is assumed to be star formation dominated). Molecular line observations in the local Universe have shown that on average galaxy mergers tend to have higher $\Lfir/\MHtwo$ ratios ($\gtrsim 100\,\Lsun\,\Msun^{-1}$) than non- or weakly-interacting star-forming galaxies \citep[e.g.,][]{Young86,Solomon88}. In addition, their far-infrared spectral energy distributions are considerably warmer \citep[IRAS $S_{60\,\micron}/S_{100\,\micron}$ colors higher than 0.6;][]{Mazzarella91}, probably indicating a dust temperature increase due to the intense UV radiation fields associated with the high concentration of young stars.

The effects of merger-induced versus quiescent star formation should also be reflected in the emission line properties of the star-forming gas. For example, the intensities of some fine structure lines produced in HII regions and photodissociation regions (PDRs) are strongly dependent on the value of the ionization parameter at the surface of the clouds \citep[e.g.,][]{Abel05}, where the ionization parameter $U$ is defined as the number of hydrogen ionizing photons ($h \nu > 13.6$\,eV) per hydrogen particle. As a first order approximation $U$ should be proportional to $\Lfir/\MHtwo$ because the number of massive stars (and hydrogen ionizing photons) per unit of molecular gas mass increases with this ratio. High ionization parameter values have been proposed to explain the low [\ion{C}{2}]\,158\,\micron\ line to far-infrared continuum (FIR\footnote{We use the definition of the far-infrared flux, FIR(42.5\,\micron--122.5\,\micron), and the far-infrared luminosity, \Lfir(40\,\micron--500\,\micron), given in table~1 of \citet{Sanders96}.}) ratios observed with the ISO satellite in some luminous and ultraluminous infrared galaxies \citep[LIRGs and ULIRGs; e.g.,][]{Luhman03,Abel09}. There are however other possible explanations to this ``[\ion{C}{2}] line deficit" (saturation of the [\ion{C}{2}] line, optical depth effects, soft UV fields; see \citealt{Luhman03} for a detailed description), that could not be fully excluded in the past due to the small number of infrared galaxies detected in far-infrared fine structure lines other than [\ion{C}{2}].

Here we report initial results from the far-infrared fine structure line observations of a sample of local galaxies obtained with \emph{Herschel} \citep{Pilbratt10}. We find that galaxies with $\Lfir/\MHtwo \gtrsim 80\,\Lsun\,\Msun^{-1}$ tend to have lower line to FIR continuum ratios than galaxies with lower $\Lfir/\MHtwo$ values. These line deficits affect all the observed lines, independent of their origin in the ionized or neutral phase of the interstellar medium.

\section{Observations and Data Reduction}

The observations discussed in this Letter are part of the \emph{Herschel} guaranteed time key program SHINING (P.I. E.~Sturm), devoted to study the far-infrared properties of a sample of more than 100 galaxies, that includes local starbursts, Seyfert galaxies, low-metallicity systems and infrared luminous galaxies at low and high redshift. Here we present the fine structure line data for the first 44 galaxies: 16 Seyferts (\object{NGC 1068} and \object{Cen A}, among others), 5 HII galaxies (e.g., \object{M 82}, \object{M 83}, \object{NGC 253}), 21 LIRGs and ULIRGs (e.g., \object{NGC 4418}, \object{Arp 220}, \object{Mrk 231}) and two high redshift galaxies (\object{IRAS F10214+4724} and \object{MIPS J142824.0+352619}). 

The observations were obtained with the PACS far-infrared spectrometer \citep{Poglitsch10} between October 2009 and December 2010. We observed the three PDR lines [\ion{C}{2}]\,158\,\micron, [\ion{O}{1}]\,145\,\micron\ and [\ion{O}{1}]\,63\,\micron, and the four HII lines [\ion{N}{2}]\,122\,\micron, [\ion{O}{3}]\,88\,\micron, [\ion{N}{3}]\,57\,\micron\ and [\ion{O}{3}]\,52\,\micron\ in most galaxies, with some exceptions due to data acquisition problems and scheduling constraints. The rest-frame wavelength of the [\ion{O}{3}]\,52\,\micron\ line is at the low sensitivity edge of the PACS range and was observed only in the high-redshift galaxies \citep[see][]{Sturm10}.   

The data were reduced using the standard PACS reduction and calibration pipeline included in HIPE 5.0. However, for the final calibration we normalized the spectra to the telescope flux and recalibrated it with a reference telescope spectrum obtained from dedicated Neptune observations. For unresolved sources, this method gives excellent agreement (differences $\leq$15\%) between the IRAS 60\,\micron\ and 100\,\micron\ flux densities and those measured with PACS when extrapolating from the continuum close to the lines. 

Most of the targeted lines were detected. For the extended galaxies, we integrated the line fluxes and FIR continuum within the 47\arcsec$\times$47\arcsec\ PACS field of view. The detailed spatial information for these sources will be discussed in a future paper.

\section{Results}

\begin{figure*}
\centering
\scalebox{0.428}{\includegraphics{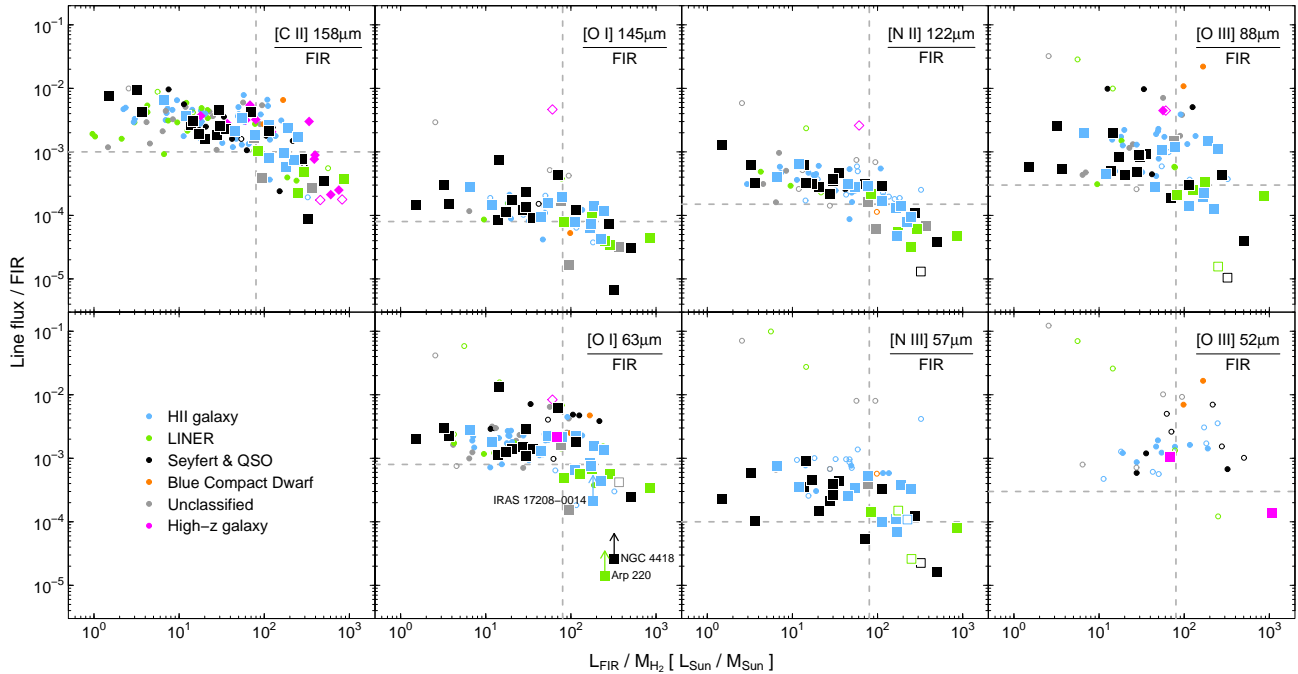}}
\caption{Global line to FIR continuum ratios in galaxies with different optical activity classifications as a function of $\Lfir/\MHtwo$. PACS observations are represented with squares. Open symbols indicate 3$\sigma$ upper limits to the line flux. All the line to FIR ratios start to decrease at $\Lfir/\MHtwo \sim 80\,\Lsun\,\Msun^{-1}$ (vertical lines). The horizontal lines indicate the approximate minimum values for the different line to FIR ratios observed in galaxies with $\Lfir/\MHtwo < 80\,\Lsun\,\Msun^{-1}$. \label{fig2}}
\end{figure*}

\subsection{The [\ion{C}{2}] Line Deficit}

The well known [\ion{C}{2}] line deficit in infrared luminous galaxies is shown in Figure~\ref{fig1}a. Together with the PACS data, we have plotted a comparison sample composed of all galaxies observed with ISO \citep{Negishi01,Malhotra01,Luhman03,Brauher08}. High redshift galaxies ($z = 1.1$--6.4) with [\ion{C}{2}] observations are also included \citep[and references therein]{Maiolino09,Hailey-Dunsheath10,Ivison10,Wagg10,Stacey10}. The average [\ion{C}{2}]$/$FIR ratio is approximately constant in galaxies with low infrared luminosities, but starts to decrease at $\Lfir \sim 10^{11}\,\Lsun$. At the same time, the scatter in the ratio increases by almost a factor of two: some high redshift galaxies do not show a [\ion{C}{2}] deficit, even if their far-infrared luminosities are higher than $10^{12}\,\Lsun$, and the source with the lowest [\ion{C}{2}]$/$FIR ratio is \object{NGC 4418}, a galaxy with $\Lfir \simeq 10^{11}\,\Lsun$. 

The scatter at high \Lfir\ can be reduced by a factor $\sim$1.5 if \Lfir$/$\MHtwo\ is used\footnote{\MHtwo\ was estimated using previously published CO observations of the galaxies. We used a $X_{\mathrm{CO}}$ conversion factor that is a continuous function of the IRAS $S_{60\,\micron}/S_{100\,\micron}$ color in a way that produces adequate values for Milky-way type galaxies ($\sim$4\,\Msun\,${L^{\scriptscriptstyle\prime}}^{-1}$, with $L^{\scriptscriptstyle\prime}$ in units of K\,km\,s$^{-1}$\,pc$^{2}$) and ULIRGs ($\sim$0.8\,\Msun\,${L^{\scriptscriptstyle\prime}}^{-1}$). Some high redshift galaxies did not have enough photometric data points to estimate their rest-frame $S_{60\,\micron}/S_{100\,\micron}$ color; $X_{\mathrm{CO}} = 2$\,\Msun\,${L^{\scriptscriptstyle\prime}}^{-1}$ was used in those cases.} instead of \Lfir\ (Figure~\ref{fig1}b), presumably because this ratio is more closely related to the properties of the clouds (gas density and dust temperature) than \Lfir\ alone \citep[e.g.,][]{Gao04}. Galaxies with similar \Lfir\ luminosities can have very different $\Lfir/\MHtwo$ ratios and the excitation conditions in their interstellar medium will also differ \citep{Gonzalez08}. With this new representation, high redshift galaxies overlap with their low redshift counterparts and \object{NGC 4418} is shifted to the right of the figure due to its relatively low molecular gas content. A similar result was obtained by \citet{Malhotra01} using the IRAS $S_{60\,\micron}/S_{100\,\micron}$ color in a smaller sample of local galaxies observed with ISO.

\subsection{A General Far-infrared Line Deficit}

Figure~\ref{fig2} shows the line to FIR continuum ratio against $\Lfir/\MHtwo$ for all the fine structure lines observed with PACS. Our main result is that we now find line deficits in both the PDR and the HII lines. All the line to FIR ratios start to decrease at $\Lfir/\MHtwo \sim 80\,\Lsun\,\Msun^{-1}$. This result also does not seem to depend on the optical activity classification of the galaxies. We note, however, that the integrated emission of the [\ion{O}{1}] and [\ion{O}{3}] lines in some Seyfert galaxies might have a significant contribution from X-ray dominated regions close to the active galactic nucleus \citep[AGN; e.g.,][]{Maloney96}. 

The [\ion{O}{1}]\,63\,\micron\ line was detected in absorption at some velocities in \object{NGC 4418}, \object{Arp 220} and \object{IRAS 17208-0014}. The [\ion{O}{1}]\,63\,\micron$/$FIR ratios used in Figure~\ref{fig2} correspond to the emission components only and should be regarded as lower limits. This will reduce the deficit observed for this line to some extent, but will not eliminate it, as other galaxies \citep[e.g., \object{Mrk 231};][]{Fischer10} with no clear absorption features in their line profiles do still have lower [\ion{O}{1}]\,63\,\micron$/$FIR ratios than galaxies with $\Lfir/\MHtwo < 80\,\Lsun\,\Msun^{-1}$. The [\ion{O}{1}]\,145\,\micron\ line is less likely to be affected by this kind of absorption and also shows a deficit at high $\Lfir/\MHtwo$. There is no sign of absorption by foreground material in the other fine structure lines at the current PACS velocity resolution ($\sim$100--250\,km\,s$^{-1}$).

In summary, our observations show that the line to FIR continuum ratio drops by a factor of 3 to 10 above $\Lfir/\MHtwo \sim 80\,\Lsun\,\Msun^{-1}$ in all the far-infrared fine structure lines we have observed. This drop seems to be a universal feature of galaxies at different redshifts and with different optical activity classifications. Neutral and ionized gas tracers are affected, as are lines with different rest-frame wavelengths, intrinsic optical depths and critical densities. This eliminates most of the proposed explanations for the [\ion{C}{2}] deficit, leaving the dependence on the $\Lfir/\MHtwo$ ratio as the most plausible dominant factor driving the deficits. We have assumed in this paper that $\Lfir$ is mostly due to star formation. This might not be true for some galaxies like \object{Mrk 231}, but even an AGN contribution of 50$\%$ to $\Lfir$ will fall short of explaining the factor of 10 decrease in some of the line to FIR ratios.

In the next section we show with quantitative modeling that high ionization parameters can indeed account for most of the observed line deficits. The key factor is that in dusty star forming regions the fraction of luminosity absorbed by dust particles in ionized as well as neutral gas increases with $U$.

\begin{figure*}
\centering
\scalebox{0.428}{\includegraphics{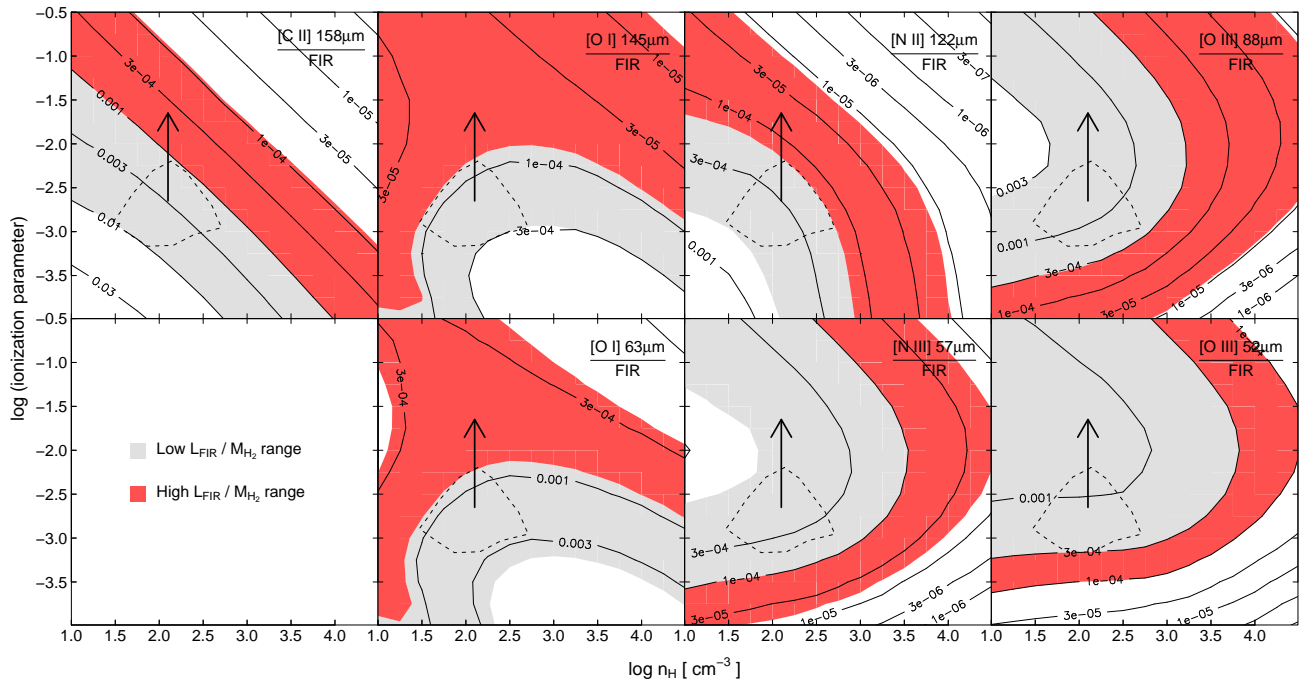}}
\caption{Cloudy model predictions for the observed line to FIR continuum ratios as a function of the hydrogen density and the ionization parameter at the illuminated face of the cloud. Red (grey) areas indicate the approximate range of values in the line to FIR continuum ratios observed in galaxies with $\Lfir/\MHtwo \geq 80\,\Lsun\,\Msun^{-1}$ ($< 80\,\Lsun\,\Msun^{-1}$). The dashed contour indicates the values of the ionization parameter and hydrogen density that can simultaneously explain all the line to FIR ratios in galaxies with $\Lfir/\MHtwo < 80\,\Lsun\,\Msun^{-1}$. The arrows show the changes to the line to continuum ratios when the ionization parameter is increased by an order of magnitude. \label{fig3}} 
\end{figure*}

\section{Combined HII and PDR Line Models}

\citet{Abel09} used the spectral synthesis code Cloudy \citep{Ferland98} to model the [\ion{C}{2}]$/$FIR ratio in clouds exposed to different radiation fields. They found that they could reproduce the [\ion{C}{2}] line deficits observed in some local LIRGs and ULIRGs, as well as their warmer IRAS $S_{60\,\micron}/S_{100\,\micron}$ colors, by increasing the value of the ionization parameter at the surface of the clouds. As $U$ increases, the HII region is extended to higher $A_{V}$ into the cloud. As a result, a larger fraction of the UV photons are absorbed by the dust in the ionized region and reemitted in the form of infrared emission. The net effect is that the fraction of UV photons available to ionize and excite the gas is reduced at high $U$, decreasing the relative intensity of the fine structure lines compared to the FIR continuum \citep[see also][]{Voit92}. Since the number of UV photons per dust particle increases with $U$, the dust temperature (and the $S_{60\,\micron}/S_{100\,\micron}$ color) increases.

We have extended the analysis in \citet{Abel09} to include all the fine structure lines observed with PACS. We used Cloudy to simultaneously calculate the emission from the molecular, neutral and ionized gas assuming pressure equilibrium in the clouds. The ionizing source was represented by a non-LTE CoStar stellar atmosphere \citep{Schaerer97} with $T_{*} = 36000$\,K, and the calculations were stopped at $A_{V} = 100$\,mag ($N_{H} = 2 \times 10^{23}$\,cm$^{-2}$ for $A_{V}/N_{H} = 5 \times 10^{-22}$\,mag\,cm$^{2}$). To facilitate the comparison with previous studies \citep{Kaufman99,Abel09}, we used the gas-phase elemental abundances from \citet{Savage96}: C$/$H $= 1.4 \times 10^{-4}$, O$/$H $= 3.2 \times 10^{-4}$, N$/$H $= 7.9 \times 10^{-5}$. In our model the spectral energy distribution of the ionization source is fixed. Thus, for a given density, an increase of the ionization parameter results in a proportional increase in the number of UV photons at all energies.

In Figure~\ref{fig3} we show the model predictions for the line ratios represented in Figure~\ref{fig2}. Models with $U \sim 10^{-3}$--$10^{-2.5}$ and hydrogen densities $\sim$10$^{1.5}$--$10^{2.5}$\,cm$^{-3}$ can simultaneously explain the average line to FIR ratios measured in galaxies with low $\Lfir/\MHtwo$ values. The observed [\ion{C}{2}], [\ion{N}{2}] and [\ion{O}{1}] line deficits in galaxies with $\Lfir/\MHtwo > 80\,\Lsun\,\Msun^{-1}$ can be easily understood if their ionization parameters are at least an order of magnitude higher, consistent with the idea that the ionization parameter is approximately proportional to $\Lfir/\MHtwo$. However, we cannot explain with this particular model the observed line deficits in the [\ion{O}{3}] and [\ion{N}{3}] high ionization lines. These lines are produced closer to the ionizing source, where the effect of the dust opacity on the formation of the lines is less important. This may indicate that the ionization parameter is not the only property that varies as $\Lfir/\MHtwo$ increases. In future papers we will explore the full parameter space in these models, varying the ionization source (SED), the stopping criteria and the metallicity of the gas. We will also study the possibility of multiple components contributing to the global far-infrared spectrum of these galaxies, including partial covering of the high ionization regions \citep[e.g.,][]{Fischer10}.

\section{Discussion and Conclusions}

As discussed in the introduction, galaxies can experience short periods of intense star formation, normally associated with the merging of massive galaxies, in which the molecular gas is transformed into stars with high efficiency. A good way to investigate this enhancement of the star formation efficiency is to study the relation between the molecular gas surface density, $\Sigma_{\mathrm{H_{2}}}$, and the star formation rate surface density, $\Sigma_{\mathrm{SFR}}$. Recent studies of the global infrared continuum and molecular gas properties of galaxies in the local and high redshift Universe suggest that, independent of redshift, luminous mergers and non- or weakly-interacting star-forming galaxies follow two separate Kennicutt-Schmidt relations ($\Sigma_{\mathrm{SFR}} \propto \Sigma_{\mathrm{H_{2}}}^N$) with similar exponents $N \simeq 1.1$--1.2, but different normalizations \citep{Genzel10,Daddi10}. The explanation for this offset in the star formation relations is still not clear, but it may indicate that, in addition to $\Sigma_{\mathrm{H_{2}}}$, other galaxy properties, like the dynamical time scale, may play a significant role regulating the star formation efficiency of the gas.

The $\Lfir/\MHtwo$ value that separates galaxies into these two types of star formation ($\sim$100\,\Lsun\,\Msun$^{-1}$) is remarkably similar to the point where we start to find a decline in the fine structure line to FIR ratios. This indicates that the average properties of the neutral and ionized gas are significantly different in these galaxies, broadly consistent with the scenario of a highly compressed more efficient star formation, creating largely enhanced ionization parameters that manifest themselves in lower line to continuum ratios. Our detection of a universal drop of the line to continuum ratio at about the same $\Lfir/\MHtwo$ value where \citet{Genzel10} and \citet{Daddi10} claim a transition to a more efficient star formation mode gives in turn credence to this interpretation. 

In the next two years \emph{Herschel} will increase the number of galaxies with fine structure line observations in the far-infrared, especially in objects with high $\Lfir/\MHtwo$. The detailed study of the line ratios, combined with the use of spectral synthesis codes like Cloudy, will provide valuable information about the ionizing sources, possible AGN contribution to \Lfir, far-infrared optical depth, and the physical properties and metallicity of the gas. This information will be critical to understand the driving mechanisms involved in these modes of star formation.

\acknowledgments

We would like to thank the referee, Greg Bothun, for his comments and suggestions, which helped to clarify the paper. PACS has been developed by a consortium of institutes led by MPE (Germany) and including UVIE (Austria); KU Leuven, CSL, IMEC (Belgium); CEA, LAM (France); MPIA (Germany); INAF-IFSI/OAA/OAP/OAT, LENS, SISSA (Italy); IAC (Spain). This development has been supported by the funding agencies BMVIT (Austria), ESA-PRODEX (Belgium), CEA/CNES (France), DLR (Germany), ASI/INAF (Italy), and CICYT/MCYT (Spain). Basic research in IR astronomy at NRL is funded by the US ONR;  J.F. also acknowledges support from the NHSC. E.G-A is a Research Associate at the Harvard-Smithsonian Center for Astrophysics.

Facilities: \facility{Herschel(PACS)}, \facility{ISO}, \facility{IRAS}

\end{document}